\documentclass{pasj01}
\Received{2017 November 21}
\Accepted{2018 February 1}
\Published{$\langle$publication date$\rangle$}

\begin{document}

\title{Usage of \texttt{pasj01.cls}}
\title{ALMA  [C\emissiontype{I}] observations toward the central region of a Seyfert galaxy NGC~613}
\author{Yusuke \textsc{Miyamoto}~\altaffilmark{1},
Masumichi \textsc{Seta}~\altaffilmark{2}, 
Naomasa \textsc{Nakai}~\altaffilmark{3, 4},
Yoshimasa \textsc{Watanabe}~\altaffilmark{3, 4}
Dragan \textsc{Salak}~\altaffilmark{2} and 
Shun \textsc{Ishii}~\altaffilmark{5, 6}
}%
\altaffiltext{1}{Nobeyama Radio Observatory, NAOJ, Nobeyama, Minamimaki, Minamisaku, Nagano 384-1305, Japan}
\email{miyamoto.yusuke@nao.ac.jp}
\altaffiltext{2}{Department of Physics, School of Science and Technology, Kwansei Gakuin University, 2-1 Gakuen, Sanda, Hyogo 669-1337, Japan}
\altaffiltext{3}{Division of Physics, Faculty of Pure and Applied Sciences, University of Tsukuba, Tsukuba, Ibaraki 305-8571, Japan}
\altaffiltext{4}{Tomonaga Center for the History of the Universe, University of Tsukuba, Tsukuba, Ibaraki 305-8571, Japan}
\altaffiltext{5}{Chile Observatory, NAOJ, 2-21-1 Osawa, Mitaka, Tokyo 181-8588, Japan}
\altaffiltext{6}{Joint ALMA Observatory (JAO), Alonso de C\'{o}rdova 3107, Vitacura 763-0355, Santiago, Chile}


\maketitle

\begin{abstract}
We report ALMA observations of  [C\emissiontype{I}]($^3P_1-^3P_0$),  \atom{C}{}{13}O, and C\atom{O}{}{18}($J=1-0$) toward the central region  of a nearby Seyfert galaxy NGC~613.
The very high resolutions of $0\farcs26\times0\farcs23$ (=$22\times20$~pc) for [C\emissiontype{I}] and $0\farcs42\times0\farcs35$  (=$36\times30$~pc) for  \atom{C}{}{13}O, and C\atom{O}{}{18} resolve the circum-nuclear disk (CND) and  star-forming ring.
The distribution of  [C\emissiontype{I}] in the ring resembles that of the CO emission, 
although  [C\emissiontype{I}]  is prominent in the CND.
This can be caused by the low intensities of the CO isotopes due to the low optical depths under the high temperature in the CND.

We found that the intensity ratios of [C\emissiontype{I}] to \atom{C}{}{12}O(3--2) ($R_{\rm C\emissiontype{I}/CO}$) and to \atom{C}{}{13}O(1--0) ($R_{\rm C\emissiontype{I}/\atom{C}{}{13}O}$) are high at several positions around the edge of the ring. 
The spectral profiles of CO lines mostly correspond each other in the  spots of the ring and high $R_{\rm C\emissiontype{I}/CO}$, but those of [C\emissiontype{I}] at spots of high $R_{\rm C\emissiontype{I}/CO}$ are different from CO.
These results indicate that [C\emissiontype{I}] at the high $R_{\rm C\emissiontype{I}/CO}$  traces different gas from that traced by the CO lines .
The [C\emissiontype{I}]  kinematics along the minor axis of NGC~613 could be interpreted as a bubbly molecular outflow.
The outflow rate of molecular gas is  higher than star formation rate in the CND.
The flow could be  mainly boosted by the AGN through its radio jets.

\end{abstract}

\section{Introduction}
Molecular hydrogen (H$_2$) 
is a major component of the interstellar medium in galaxies.
While $^{12}$CO($J=$1--0) line has been used as a principal probe to trace molecular gas and  to study dynamics and molecular gas distribution in galaxies, 
\atom{C}{}{12}O is not easy to estimate the gas mass due to the large optical depth.
Recently, atomic carbon (C\emissiontype{I}) attracts attention as a good tracer of mass of the cold 
molecular gas especially in high redshift galaxies or metal-poor objects
 (e.g., \cite{papado_greve2004}, \cite{papado_etal2004}).

It has been recognized that C\emissiontype{I} exists predominantly in the thin layer near the surface of molecular clouds exposed to UV radiation, which is called a photodissociation region (PDR) (\cite{tielens1985}, \cite{hollenbach1991}).
However, large-scale  C\emissiontype{I}($^3P_1-^3P_0$), hereafter simply referred as [C\emissiontype{I}],  mapping of the Orion A and B molecular clouds revealed that  [C\emissiontype{I}] coexists with \atom{C}{}{12}O and \atom{C}{}{13}O(1--0) and their intensities correlate each other \citep{ikeda2002}. 
Since the critical density of  [C\emissiontype{I}] is similar to that of \atom{C}{}{12}O ($n\approx 10^3$~cm$^{-3}$),
these findings indicate that the emission lines 
arise from the same volume and share similar excitation temperature.  

The global extent of the  [C\emissiontype{I}] emission is similar to that of dense molecular gas traced by C$^{18}$O, 
whereas the intensities of  [C\emissiontype{I}] and C$^{18}$O anti-correlates to each other (e.g., \cite{maezawa1999}).
Oka et al. (2005) showed distribution of the  
C\emissiontype{I}-to-\atom{C}{}{12}O intensity 
ratio on the Galactic scale and suggested that the locations of 
the high intensity ratio correspond to the upstream of spiral arms. 
These results indicate that  [C\emissiontype{I}] abundance is high in the early stage of chemical evolution (within a timescale for conversion from  C\emissiontype{I} to CO; $\sim10^6$~yr) and  [C\emissiontype{I}] traces young clouds which are just forming dense cores (\cite{suzuki1992}, \cite{maezawa1999}).

A couple of dozen observations toward nearby galaxy centers with single-dish telescopes whose linear resolution of several 100~pc 
have been found that the intensity ratio of [C\emissiontype{I}] to \atom{C}{}{13}O in most galaxy centers exceeds unity, 
being rare for the Galactic molecular clouds, 
and is likely related with the central activities \citep{israel}.
The central outflow or shock may enhance [C\emissiontype{I}]  compared  with CO(1--0) \citep{krips}, 
indicating a possibility of [C\emissiontype{I}] as an outflow or shock tracer.
On the other hand, 
the distribution of [C\emissiontype{I}] is still unclear
due to a few mapping observations with the limited spatial resolution (e.g., \cite{gerin}).
For characterizing  molecular clouds traced by C\emissiontype{I}, 
it is necessary  to clarify the distributions of C\emissiontype{I} and 
CO lines
relative to energetic activities, such as star formation and central outflow.

Molecular gas in barred galaxies can be transported toward the galactic center,
because the galactic bar can effectively drain the angular momentum of the gas (\cite{binney}, and references therein).
The gas gathers at the region of nearly circular orbits (x$_2$ orbits) and forms a nuclear ring which is the site of vigorous star formation ($ r \sim$ a few 100 pc; we refer to a star-forming ring).
Star-forming rings are promising reservoirs of the gas that feeds to the accretion disks in active galactic nuclei (AGN) via the circum-nuclear disk (CND; $r\sim1-100$~pc).

The nearby galaxy NGC~613 hosts a low-luminosity AGN and  prominent radio jets from the center (e.g., \cite{goulding}, \cite{hummel1992}). 
The central  CND of NGC~613 ($r\lesssim90$~pc) contains abundant molecular gas, 
while the star formation rate (SFR) in the CND is lower than that in its star-forming ring ($250\lesssim r \lesssim 340$~pc),
which is probably caused by the interaction between the jets and gas in the CND (\cite{falcon}, \cite{davies}, \cite{miyamoto}).
Comparisons between  [C\emissiontype{I}] and 
CO lines
in the energetic activities could provide clues to 
understanding what [C\emissiontype{I}] traces.
This letter reports [C\emissiontype{I}] and CO observations with higher resolution enough to resolve the central region of the galaxy, i.e., CND and star-forming ring, for the first time as a galaxy.

\section{Observations}
\label{sec:observation}
\begin{figure*}[h]
 \begin{center}
  \includegraphics[width=0.8\linewidth]{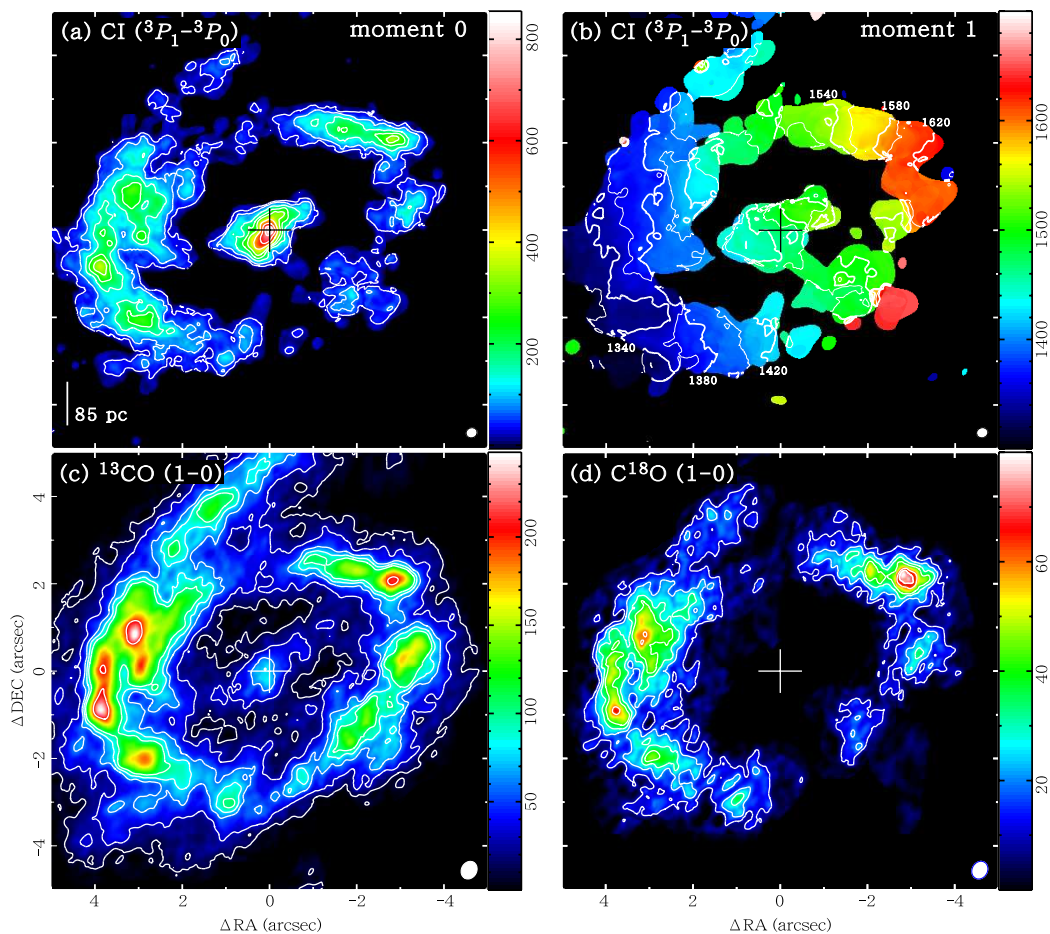}
 \end{center}
 \caption{
(a) Integrated intensity map of C\emissiontype{I}($^3P_1$--$^3P_0$). The contours are 5,10,20,30,40,50,60,70 and 80 $\sigma$ where 1$\sigma=10$~K~km~s$^{-1}$.
(b) The [C\emissiontype{I}]  velocity field map derived from intensity-weighted mean velocities. 
The velocity is in km~s$^{-1}$ with respect to LSR and the radio definition.
(c) Integrated intensity map of \atom{C}{}{13}O($J=1-0$). 
The contours are 5, 15, 30, 40, 50, and 100 $\sigma$ where 1$\sigma=2$~K~km~s$^{-1}$.
(d) Integrated intensity map of C\atom{O}{}{18}($J=1-0$). 
The contours are 5, 10, 15, 20, 35, 30, 35, 40, and  45 $\sigma$ where 1$\sigma=2$~K~km~s$^{-1}$.
A cross in each map is the peak position of the continuum emission, i.e., the galactic center, 
$(\alpha_{\rm J2000.0}$, $\delta_{\rm J2000.0})=$(\timeform{1h34m18.19s}, \timeform{-29D25'06.60''}).
The filled ellipse in the bottom right corner of each map represents the beam size, 
$0\farcs26\times0\farcs23$ ($\theta=\timeform{-71D}$) for  [C\emissiontype{I}] and 
 $0\farcs42\times0\farcs35$ ($\theta=\timeform{-29D}$) for  \atom{C}{}{13}O and C\atom{O}{}{18}($1-0$).
}
 \label{fig:map}
\end{figure*}

NGC~613 was observed with the Atacama Large Millimeter/submillimeter Array (ALMA) using Bands~3 and 8 receivers.
For Band~8, the Atacama compact array (ACA) and total power array (TP) were used in addition to the 12-m array.
The synthesized beams at Bands~3 and 8 were $0\farcs42\times0\farcs35$ ($\theta=\timeform{-29D}$) and $0\farcs26\times0\farcs23$ ($\theta=\timeform{-71D}$), 
corresponding to $36\times30$ and $22\times20$~pc, respectively, at the distance of the galaxy (17.5 Mpc; \cite{tully1988}).
Single-point observations and three-point mosaic observations 
were conducted for Band~3 and Band~8, respectively. 
The maximum recoverable scale for Band~3 is $\sim \timeform{22''}$. 
These setups allowed us to image the CND  and star-forming ring of NGC~613. 
The phase reference center of $(\alpha_{\rm J2000.0}$, $\delta_{\rm J2000.0})=$
(\timeform{1h34m18.19s}, \timeform{-29D25'06.60''}) was adopted (Miyamoto et al. 2017). 
The correlators for Band~3 were configured to set three spectral windows 
in the upper sideband to measure \atom{C}{}{13}O(1--0) ($\nu_{\rm rest}=$110.201354~GHz) and C$^{18}$O(1--0) ($\nu_{\rm rest}=$109.782176~GHz) in the 2SB dual-polarization mode.
The correlators for Band~8 were configured to set one spectral window in the upper sideband  to cover  C\emissiontype{I}($^3P_1-^3P_0$) ($\nu_{\rm rest}=$492.160651~GHz), 
in the dual-polarization mode.
The flux density at the Band~3 was calibrated by J2357-5311 and J0334-4008, and at Band~8 by J0006-0623 and Uranus.
The time variations of amplitude and phase were calibrated by J0106-2718 and J0145-2733 for Band~3 and by J0204-1701 and J0137-2430 for Band~8.

The data were processed using the Common Astronomy Software Application (CASA; \cite{mcmullin2007}). 
The velocity resolution of each line data obtained at different observing tracks with the 12~m array and ACA
were separately rearranged  to be 10~km~s$^{-1}$. 
Each line data was then combined after subtracting continuum emission determined at the channels that are free from spectral line emission. 
To image the continuum emission, 
we used the flux density at the emission-free channels.
The imaging was performed using the CLEAN-algorithm in CASA.
CLEAN maps were obtained considering the Briggs weighting mode on the data with robustness of 0.5 for Band~3 and the natural weighting mode for Band~8. 
The resultant maps were $1500\times1500$~pixels with $0\farcs05$~per pixel and  $0\farcs02$~per pixel for Band~3 and 8, respectively. 
For line emission of Band~8, 
TP data were calibrated through flagging and the system temperature correction,
and imaged independently from the 12-m array and ACA data. 
By using the Feather algorithm in CASA, the low-resolution image obtained through the TP and high-resolution image obtained through the 12-m and ACA were converted  into the gridded visibility plane and combined. 
Finally, the data was reconverted into a combined image.
The sensitivities in the resultant cube of  [C\emissiontype{I}], \atom{C}{}{13}O, and  C\atom{O}{}{18}($J=1-0$) are 6~mJy~beam$^{-1}$
($\sim0.5$~K),  
0.2~mJy~beam$^{-1}$
($\sim0.1$~K),  
and 0.2~mJy~beam$^{-1}$
($\sim0.1$~K),  
respectively, in channels of 10~km~s$^{-1}$ width.

We found that the peak position of 100-GHz continuum 
corresponds to the phase center with the uncertainty of $\sim 0\farcs01$, 
while the peak position of 490-GHz continuum is offset from  the phase center by $\sim 0\farcs07$, 
which is caused by  large separation of $\timeform{14.2D}$ between NGC~613 and the phase calibrator (J0204-1701).
We, therefore, shifted the position in Band~8 data so that the continuum position coincides with the center.

\section{Results and Discussion}
\label{sec:result}
\subsection{Distributions of [C\emissiontype{I}]  and CO lines}

\begin{figure}
 \begin{center}
  \includegraphics[width=\linewidth]{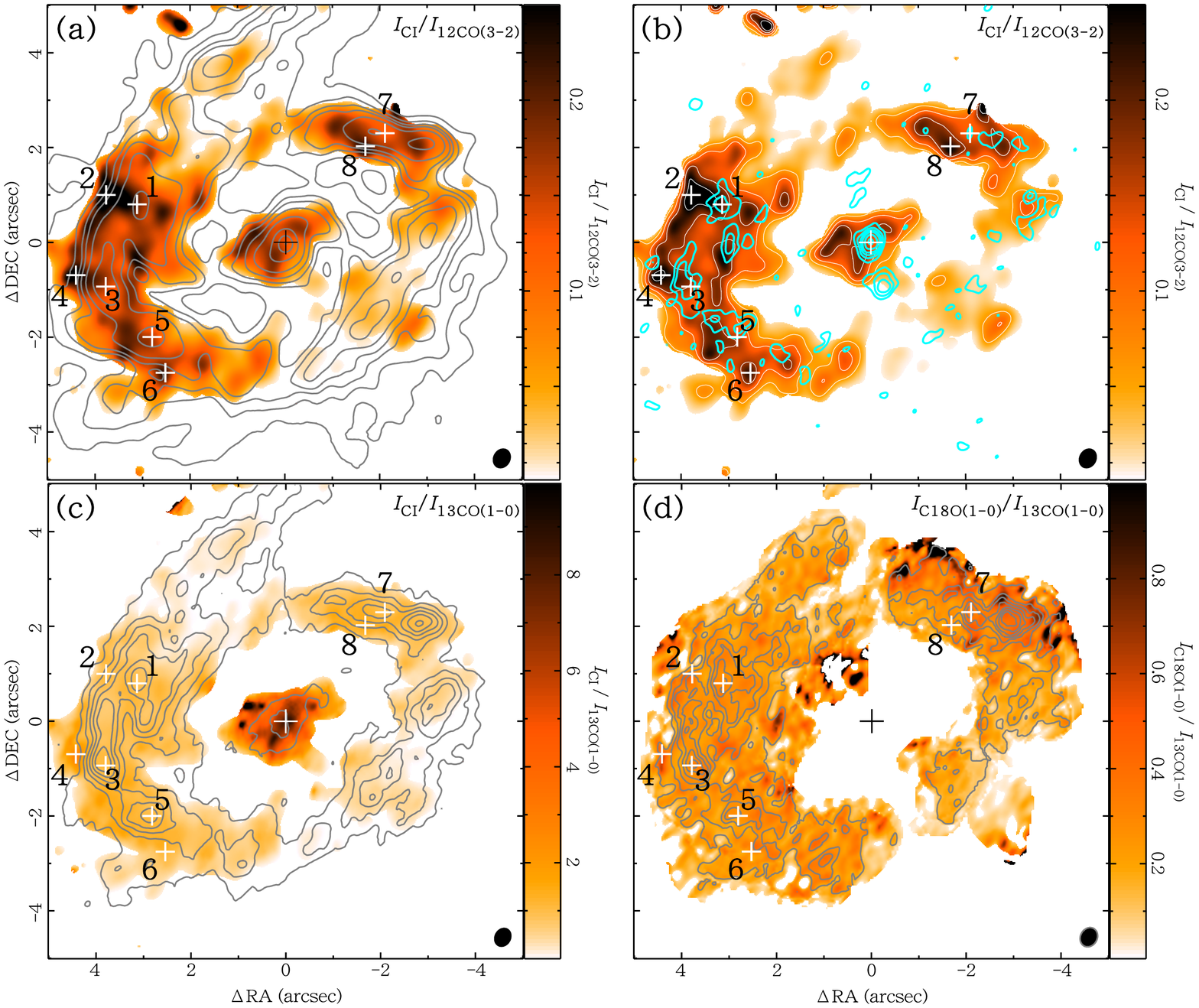}
 \end{center}
 \caption{
 (a) Map (color) of the integrated intensity ratio of $I_{\rm C\emissiontype{I}}$ to $I_{\rm CO(3-2)}$, which is from \citet{miyamoto}, overlaid with $I_{\rm CO(3-2)}$ contours of 5,10,15,20,30,50,100,150 and 180 $\sigma$ where 1$\sigma=20$~K~km~s$^{-1}$. 
 (b) Same as (a) but contours of the 100-GHz flux (3,5,7,9,12 and 15 $\sigma$ where $\sigma=20~{\rm \mu Jy}$~beam$^{-1}$). 
 (c) Map (color) of the integrated intensity ratio of $I_{\rm C\emissiontype{I}}$ to $I_{\rm \atom{C}{}{13}O(1-0)}$, overlaid with  $I_{\rm \atom{C}{}{13}O(1-0)}$ contours  (same as figure~\ref{fig:map}(b)). 
 (d) Map (color) of the integrated intensity ratio of $I_{\rm C\atom{O}{}{18}(1-0)}$ to $I_{\rm \atom{C}{}{13}O(1-0)}$, 
overlaid with  $I_{\rm  C\atom{O}{}{18}(1-0)}$ contours  (same as figure~\ref{fig:map}(c)). 
}
 \label{fig:ratio}
\end{figure}

Figures~\ref{fig:map}(a), (c), and (d) show the integrated intensity maps of  [C\emissiontype{I}], \atom{C}{}{13}O($J=$~1--0), and C\atom{O}{}{18}($J=$~1--0), respectively, 
where pixel values $<3~\sigma $ in each velocity channel of the cube smoothed with twice the size of the beam-width were masked 
to derive the integrated intensities of the weak emission accurately and enhance the contrasts, and figure~\ref{fig:map}(b) shows  the intensity-weighted [C\emissiontype{I}] velocity field map. 
The [C\emissiontype{I}] line emission is detected both in the star-forming ring and CND,
while C\atom{O}{}{18} is faint in the CND. 
The velocity field of [C\emissiontype{I}] is consistent with that of \atom{C}{}{12}O(3--2) shown by \citet{miyamoto}, i.e., rigid rotation in the ring.
The peak positions and distribution of the [C\emissiontype{I}] emission in the ring are  consistent with those of the CO emission, although being inconsistent with them in the CND.
According to \citet{ikeda2002}, 
we calculated the optical depths of [C\emissiontype{I}], \atom{C}{}{13}O(1--0) and C\atom{O}{}{18}(1--0) under LTE assumption, 
where those excitation temperatures are similar, 
and we adopted the rotational temperatures of $T_{\rm rot}\sim 18$~K in the CND and $T_{\rm rot}\sim 12$~K in the ring as the excitation temperatures \citep{miyamoto}. 
The optical depths of  [C\emissiontype{I}] in the CND and in the ring are estimated to be 
$\tau_{\rm  C\emissiontype{I}}= 0.6\pm0.3$ 
and $\sim 0.1-1.5$, respectively, 
and those of \atom{C}{}{13}O(1--0) and C\atom{O}{}{18}(1--0) are thin both in the CND and ring; 
$\tau_{\rm \atom{C}{}{13}O}= 0.06\pm0.01$ (CND) 
and $\sim 0.1-0.4$ (ring),  
and $\tau_{\rm C\atom{O}{}{18}}\sim 0.02\pm0.01$ (CND) 
and $\sim 0.02-0.08$ (ring).
The relatively high optical depth of  [C\emissiontype{I}] in the ring can be caused by the underestimation of the  excitation temperature. 
The corresponding distributions of [C\emissiontype{I}] and \atom{C}{}{13}O(1--0) in the ring are consistent with the previous studies about the Galactic components (e.g., \cite{shimajiri}), 
indicating a clumpy PDR model, in which UV radiation can penetrate  deeper in a clumpy cloud \citep{spaans}, 
and/or  chemical evolution model (e.g., \cite{oka2001}, \cite{ikeda2002}).
\begin{figure}
 \begin{center}
  \includegraphics[width=\linewidth]{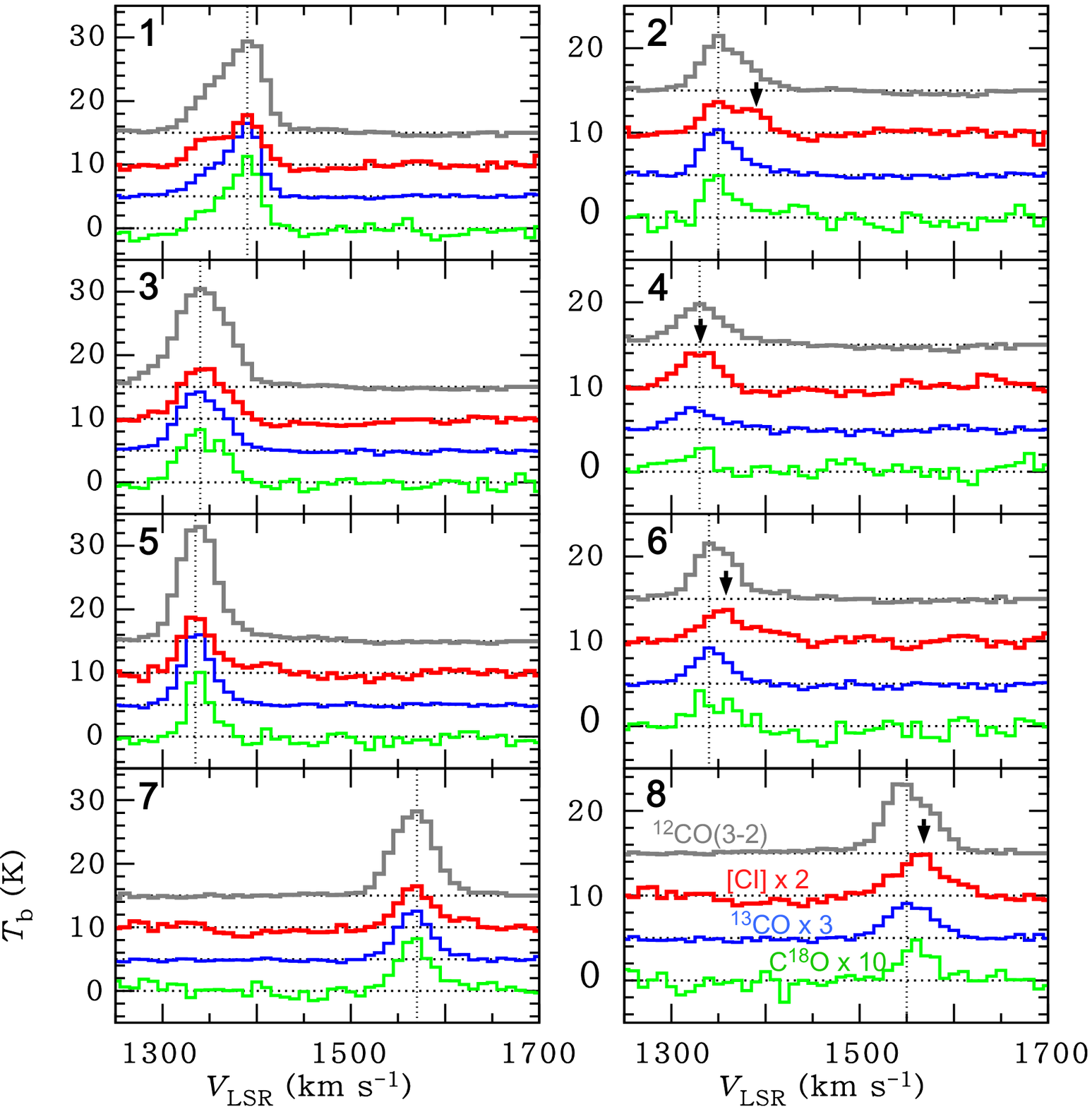}
 \end{center}
 \caption{
Spectra of [C\emissiontype{I}], CO(3-2), \atom{C}{}{13}O(1-0) and C\atom{O}{}{18}(1-0) at the spots denoted in figure~\ref{fig:ratio}. 
The corresponding aperture is 0\farcs5 (42~pc). 
 }
 \label{fig:spe}
\end{figure}

\begin{figure*}
 \begin{center}
  \includegraphics[width=0.9\linewidth]{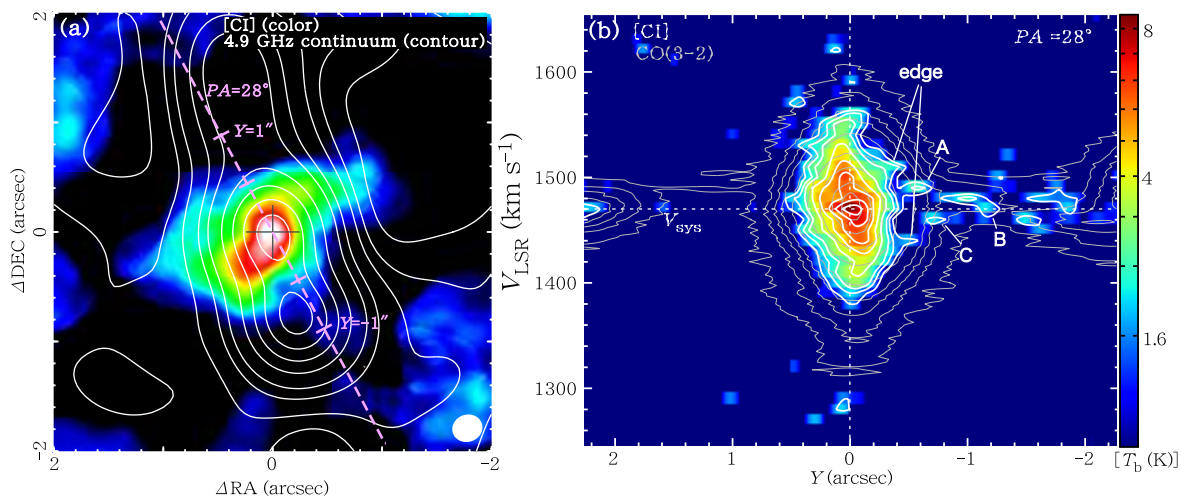}
 \end{center}
 \caption{
(a) Enlarged view of the integrated intensity maps of  [C\emissiontype{I}]  in figure~\ref{fig:map}(a) (color) is overlaid with the 4.9 GHz continuum from \citet{miyamoto} (contours).
(b) Position-velocity (PV) diagram of [C\emissiontype{I}] along the minor axis of NGC~613; ${\it PA}=\timeform{28D}$ \citep{miyamoto} (Color image and white contours).
The levels of the contours are  
3, 4, 5, 7, 9, 11, 13, 15, and 17~$\sigma$ where $\sigma=0.5~{\rm K}$.
Gray contours show the PV diagram of CO(3--2) from \citet{miyamoto}, 
and the levels are 5, 15, 20, 30, 50, 70, 90, and 100~$\sigma$ where $\sigma=0.25~{\rm K}$.
 }
 \label{fig:out}
\end{figure*}

Figure~\ref{fig:ratio}(a) and (c)  show the integrated line ratios of [C\emissiontype{I}] to \atom{C}{}{12}O(3--2) ($\equiv R_{\rm CI/CO}$) and to  \atom{C}{}{13}O(1--0) ($\equiv R_{\rm CI/\atom{C}{}{13}O}$),  
where the \atom{C}{}{12}O intensity map is from \citet{miyamoto} and the beam size of [C\emissiontype{I}] is convolved with those of the CO.
For comparison, 
the distribution of 100-GHz continuum, whose flux density can be dominated by free-free thermal emission from H\emissiontype{II} regions followed by non-thermal emission,  
such as radio jets (e.g., \cite{condon}, \cite{salak}),
superposed on the ratio map of $R_{\rm C\emissiontype{I}/CO}$ in figure~\ref{fig:ratio}(b).
The $R_{\rm C\emissiontype{I}/CO}$ is  in a range of 0.1--0.3 in both the CND and star-forming ring and there are several high  $R_{\rm C\emissiontype{I}/CO}$ spots
at the edge of the ring.
The temperature ratio of [C\emissiontype{I}] to CO at the spots of the edge, $T_{\rm C\emissiontype{I}}/T_{\rm CO}\sim0.6$,  is  enhanced more than double that in the ring, $\sim0.1-0.3$ (see channel map in supplementary material). 
The high $R_{\rm C\emissiontype{I}/CO}$ likely locates around the bright spots in 100-GHz continuum, although not being always, e.g., south-west region.
The $R_{\rm C\emissiontype{I}/\atom{C}{}{13}O}$ is strikingly very high in the CND due to the low intensity of \atom{C}{}{13}O, 
although being high at the edge of the ring as the case in $R_{\rm C\emissiontype{I}/CO}$.
The low intensity of \atom{C}{}{13}O relative to \atom{C}{}{12}O (and C\emissiontype{I}) in the CND can be due to the low optical depth, 
not by the extreme abundance of \atom{C}{}{13}O, which is caused by selective photo-dissociation, fractionation, or nucleosynthesis (e.g., \cite{langer}, \cite{vandishoeck}, and \cite{casoli}). 
The selective photo-dissociation and nucleosynthesis are inconsistent with 
 the lower star formation activity in the CND  than that in the ring \citep{falcon}.
In addition,
the lower intensity ratio of C\atom{O}{}{18} to  \atom{C}{}{13}O
in the CND  than that in the ring [figure~\ref{fig:ratio}(d)] cannot be explained by the fractionation, since the high temperature in the CND (e.g., $T_{\rm k}=350-550$~K; \cite{miyamoto}) makes the formation of \atom{C}{}{13}O ineffective, by contrast with C\atom{O}{}{18} which does not undergo the fractionation.
We found that $R_{\rm C\emissiontype{I}/\atom{C}{}{13}O}$  of $\sim 10$ in the CND is consistent with the ratio of the optical depths of $\tau_{\rm C\emissiontype{I}}=0.6\pm0.3$ to $\tau_{\rm \atom{C}{}{13}O}=0.06\pm0.01$.
In addition, it is expected that the low optical depth and high temperature in the CND cause the ratio of the fractions of the upper state level of \atom{C}{}{13}O (and C\atom{O}{}{18}) to the $J=1$ level.
In  NGC~1068, a  case that the intensities of CO isotopic species of the $J=1-0$ transition in the CND are lower than those in the ring and those of the higher transition $J=3-2$ in the CND was reported (\cite{takano2014}, \cite{nakajima})

Figure~\ref{fig:spe} shows the spectra of [C\emissiontype{I}], \atom{C}{}{12}O(3--2), \atom{C}{}{13}O(1--0) and C\atom{O}{}{18}(1--0) at 
representative positions 
on the CND and  ring.
Spots~1, 3, 5, and 7 are peaks on the ring traced by CO(3-2) and \atom{C}{}{13}O(1-0) [see figure~\ref{fig:ratio}(a) and (c)], 
whereas spots~2, 4, 6, and 8 show the high intensity ratio of $R_{\rm C\emissiontype{I}/CO}$ at the edge of the ring.
The  profiles of the CO lines mostly correspond each other at all spots, 
but those of C\emissiontype{I} at spots of high $R_{\rm C\emissiontype{I}/CO}$  
have different velocity features from those of CO.
The different spectral profiles at the high $R_{\rm C\emissiontype{I}/CO}$ 
suggest that C\emissiontype{I} traces different gas from that traced by the CO lines.
In the high  $R_{\rm C\emissiontype{I}/CO}$ at the edge of the ring, [C\emissiontype{I}] would trace dark CO, e.g., an early stage of chemical evolution \citep{tanaka2011}, 
although other models including the PDR model cannot be excluded. 
In order to clarify the reason of the different spectral profiles, multi-wavelength observations  with high angular resolution enough to resolve individual molecular clouds ($\lesssim10$~pc) are needed, 
since PDR models are developed to represent the structure of an individual cloud.

\subsection{[C\emissiontype{I}]  outflow in circum nuclear disk (CND)}

Figure~\ref{fig:out}(b) shows the position-velocity (PV) diagram of [C\emissiontype{I}]  along the minor axis with ${\it PA}=\timeform{28D}$ of NGC~613 [figure~\ref{fig:out}(a)], 
superposed on the PV diagram of CO(3--2).
The gas motion in NGC~613 is counterclockwise and the southern part of NGC~613 is on the near side (figure~\ref{fig:map}(b), \cite{burbidge}).
In the southern region ($Y\sim -1\farcs0$), we found some compact components (A, B, and C) whose size of $\sim 0\farcs4(=34$~pc) and velocity width of $\sim10~{\rm km~s^{-1}}$, being consistent with those of GMC  (e.g., \cite{sanders}).
Especially, the location of component~B at $Y\sim -1\farcs0 (= 85~{\rm pc})$ is close to the peak of the southern bubble 
traced by 4.9 GHz continuum, whose flux density is dominated by synchrotron emission due to the nuclear jets \citep{hummel1992}. 
The [C\emissiontype{I}] intensities of 31.4, 41.7, and 14.7~${\rm K~km~s^{-1}}$ of components A, B, and C correspond to the column densities of $N_{\rm {\rm C\emissiontype{I}}} = 5.4$, $7.2$, and $2.5\times10^{17}~{\rm cm^{-2}}$, respectively, according to \citet{ikeda2002} and $\tau_{\rm C\emissiontype{I}}\sim0.6$ in the CND.
The ratio of $N({\rm C\emissiontype{I}})$ to $N({\rm CO})$ in the ring becomes $\sim 0.1$, 
where we adopted $N({\rm CO}) = 7.0 \times 10^{16}~\int T_{\rm b}{\rm (\atom{C}{}{13}O)}~dv ~\tau_{\rm \atom{C}{}{13}O}/[1-\exp(-\tau_{\rm \atom{C}{}{13}O})]~{\rm [cm^{-2}] }$ by assuming $T_{\rm ex}{\rm (ring)}\sim12$~K and ${\rm [\atom{C}{}{12}O]/[\atom{C}{}{13}O] =60}$ (cf. \cite{ikeda2002}).
We roughly estimated  molecular gas masses of components A, B, and C 
to be  7.7, 10.2, and $3.6\times10^4$~\MO, respectively,
with manners similar to \citet{israel2001}, 
but adopting typical value of [C]/[H]$\sim1.5\times10^{-3}$ (e.g., \cite{israel2001}, \cite{israel2003}).

The molecular clouds may expand spherically because of the symmetric velocity patterns of the components and edge relative to $V_{\rm sys}$ in figure~\ref{fig:out}(b). 
We assumed that the velocity difference of the edge to $V_{\rm sys}$ is the same as the expanding velocities of all components, 
$v_{\rm exp}\sim35 [=(1510-1440)/2] / (\cos~i_{\rm out})~{\rm km~s^{-1}}$,
where $i_{\rm out}$ is an angle between a normal line to the bubble and the line of sight.
The offset between the averaged velocity of the edge and $V_{\rm sys}$ could be caused by an inclination of the outflow to the line-of-sight,
although the velocity resolution of $\Delta V=10$~km~s$^{-1}$ is not enough to discuss further.
For the hight of the component~B of $\sim 85~{\rm pc}$ 
and 
the expanding velocity of $v_{\rm exp}\sim35~{\rm km~s^{-1}}$ as a lower limit, 
the expanding time is $t_{\rm exp}\sim2.4~{\rm Myr}$, 
and the mass outflow rate 
of only component C becomes $dM_{\rm H_2}/dt\sim 0.04$~\MO~${\rm yr^{-1}}$,
which is higher than SFR in the CND, $0.02~{\rm \MO~yr^{-1}}$ \citep{falcon}.
It is noted that the mass outflow rate in nearby starburst galaxies is comparable to SFR, e.g., M82 \citep{salak2013}, NGC~253 \citep{bolatto}, NGC~1808 \citep{salak2016}.
The AGN bolometric luminosity of NGC~613 is  a few $10^{42}~{\rm erg~s^{-1}}$, 
accounting for $\sim10~\%$ of total luminosity including contribution of shock excitation and star formation \citep{davies}.
In addition, the Eddington luminosity is $L_{\rm edd} = 1.3\times10^{45}~{\rm erg~s^{-1}}$ by adopting the expected BH mass of $\sim 10^7\MO$ \citep{beifiori}.
The kinetic energy and luminosity of the outflowing component become 
$1/2~M_{\rm H_2}~v^2_{\rm exp} \sim 1.2\times10^{51}~{\rm erg}$ 
and $1/2~dM_{\rm H_2}/dt~v^2_{\rm exp} \sim 1.7\times10^{37}~{\rm erg~s^{-1}}$, respectively.
The jet power of $P_{\rm jet}=1.5\times10^{42}~{\rm erg~s^{-1}}$, which can be estimated by adopting the central 1.4~GHz luminosity of $3.98\times10^{37}~{\rm W~Hz^{-1}}$ \citep{hummel1985} to a scaling relation suggested by \citet{cavagnolo}, is larger than the kinetic luminosity. 
The ratio of $P_{\rm jet}/L_{\rm edd}=1.2\times10^{-3}$ is consistent with a value expected from a numerical simulation, in whose range the jet can drive the interstellar medium efficiently \citep{wagner}.
These results indicate that the jet contributes driving the gas outflow, 
although the [C\emissiontype{I}] outflow is tentative.
In order to confirm the outflow, 
more sensitive [C\emissiontype{I}] observations  are needed.

\begin{ack}
This paper makes use of the following ALMA data: ADS/JAO. ALMA\#2015.1.01487.S. 
ALMA is a partnership of ESO (representing its member states), NSF (USA) and NINS (Japan), together with NRC (Canada) and NSC and ASIAA (Taiwan) and KASI (Republic of Korea), in cooperation with the Republic of Chile. The Joint ALMA Observatory is operated by ESO, AUI/NRAO and NAOJ. 
This work was financially supported by Grants-in-Aid for Scientific Research (KAKENHI) of the Japanese society for the Promotion of Science (JSPS, Grant No. 16H03961). 
\end{ack}

\end{document}